Considerations on an Optical Test of Popper's Experiment.

J. Reintjes
Sotera Defense Solutions
reintjes@ccs.nrl.navy.mil

M. Bashkansky
Naval Research Laboratory
mark.bashkansky@nrl.navy.mil

ABSTRACT

We present a detailed analysis of a previously published realization of Popper's experiment using entangled ghost imaging. Our analysis, which is based on optical diffraction integrals, shows that, for the configuration previously described, the transverse spread of an unmeasured particle (the signal photon here) does not increase in inverse proportion to the width of its virtual confinement when its entangled twin is confined in transverse dimension by a physical slit. Rather we show that the spread of the unmeasured particle carries no dependence on the width of its virtual confinement in the published configuration. Instead, it spreads geometrically at a rate determined by the numerical aperture of the ghost imaging system. We further propose an alternative configuration for which the spread of the unmeasured particle does increase in inverse proportion to the width of its virtual confinement. We discuss the relation of these results to the predictions of Popper.

INTRODUCTION

Popper [1] described a thought experiment in which one particle of an entangled pair is confined in transverse dimension by a physical slit. Popper claimed that quantum mechanics predicted that the second (unmeasured) particle would be confined to an equal transverse width and that, as a result, its transverse momentum would increase in inverse proportion to its confined width because of the Heisenberg uncertainty principle. Popper further asserted that these predictions of quantum mechanics were incorrect and that the spread (transverse momentum) of the second particle would not increase in the absence of interaction with a physical slit. Effectively, Popper predicted that it is possible to have simultaneous knowledge of the position and momentum of the second particle with an accuracy greater than that allowed by the Heisenberg uncertainty principle.

Here, we revisit the experimental realization of the modification of Popper's experiment [1] reported previously in [2] and [3]. There the authors describe a ghost imaging system using entangled signal-idler photon pairs generated via spontaneous parametric down conversion (SPDC) in a nonlinear optical crystal. They state that the transverse extent of the signal beam in the ghost image plane is confined by a correlated measurement of the idler beam after it has passed through a narrow slit, even if there is no slit in the signal beam. They present measurements of the transverse extent of the signal beam in a diffraction plane downstream from the ghost image plane which show that the signal beam spreads after the ghost image plane, but spreads even more if a physical slit is placed in the signal beam at the ghost image plane.

The authors of [2,3] maintain that the smaller spread of the signal photon that they observed when the physical slit is not present at the ghost image plane is in agreement with Popper's assessment, that is, that the position and momentum of the second particle can be known simultaneously with an accuracy greater than that allowed by conventional quantum mechanics. They then discuss this result in terms of a potential violation of the Heisenberg uncertainty principle.

In addition to the experiments in [2] and [3], the Popper concept has been discussed extensively in the literature [4 -10] and has also been discussed in relation to a separate experiment using entangled photons [11]. Published comments on the optical experiments range from explanations of the measurements in [2] and [3] as being due to spread of the signal distribution in the ghost plane due to the finite diameter of the pump [8], to statements that there is no confinement of the unmeasured particle in the ghost plane [10]. All of these comments are based on a treatment of the behavior of the signal and idler photons in analogy to standard quantum particles described by position and momentum, although a qualitative description of the system in [2,3] was given in terms of an equivalent incoherent linear optics system [8]. There has been no analysis of the system in [2] and [3] using optical diffraction integrals, although such an analysis has been given for the experiment of Ref. 11 in [12].

Here we present a detailed analysis of the system described in [2] and [3] using optical propagation equations for the entangled signal and idler photons from their source at the SPDC crystal to their respective detectors, including effects of the finite pump beam diameter. Our results are in agreement with some of the conclusions presented in the literature, but disagree with others.

When we use the parameters appropriate for the measurements in [2] and [3] our results are in reasonable agreement with the measurements presented there. In particular, we find that the distribution of the signal photons in the ghost image plane is described by a mode profile with a full width at half maximum (FWHM) that is comparable to the width of the idler slit, in agreement with the claim in [2,3]. However, we find that the wings of the mode profile of the signal in the ghost image plane extend beyond the dimension of the idler slit. Consequently, when a physical slit is placed in the signal beam, it blocks the wings of the signal mode profile and increases the spread of the signal beam in the downstream diffraction plane. As a result we conclude that the effect of the physical slit on the spread of the signal beam that was reported in [2,3] is explained entirely by the relative sizes of the slit and the beam mode profile and the laws of optical diffraction, and is unrelated to the conjecture of Popper.

We also change some of the parameters to investigate the details of the Popper concept within this geometry. When we consider a larger numerical aperture than was used in [2,3] the mode profile of the signal beam in the ghost image plane reproduces the shape of the slit much more faithfully, and the spread of the signal beam downstream of the ghost image plane is not affected by the presence of a slit at the ghost image plane, confirming our explanation of the previous experimental observation.

We derive a scaling law for the width of the signal distribution in the diffraction plane when the idler detector behind the slit collects all the transmitted idler photons. We show that this width is independent of the width of the idler slit and the wavelength. As a result, the widths of the signal distribution in the ghost image plane and in the diffraction plane are not Fourier transform pairs or quantum conjugate variables and are therefore not restricted by a Heisenberg-like relation. Rather, the width of the signal distribution increases geometrically after the ghost image plane.

Finally we analyze a geometry in which a point idler detector is used in the focal plane of a collecting lens behind the slit, detecting only a subset of the transmitted idler photons that are associated with a single k-vector of the slit. For this arrangement we show analytically and numerically that, when the larger numerical aperture is used, the signal distribution in the ghost image plane again faithfully reproduces the idler slit, but now the distribution in the diffraction plane is the Fourier transform of the amplitude of the signal in the ghost image plane. As a result, for this configuration the width of the signal distribution in the diffraction plane increases in inverse proportion to the width of the signal distribution in the ghost image plane (equal to the width of the idler slit) even if there is no physical slit in the ghost image plane.

We discuss our results in relation to the predictions of quantum theory as set forth by Popper and his assessment of their accuracy.

ENTANGLED GHOST IMAGING

The geometry for entangled ghost imaging [13, 14] is shown in Fig. 1. Signal-idler pairs are generated in a SPDC crystal pumped using a Gaussian pump beam with a radius of $a_P$ at the 1/e field amplitude point . The SPDC process is type II with collinear phasematching. The signal and idler are separated with a polarizing beam splitter. The signal propagates a distance $d_3$ and is detected with a point-like detector D2 that is scanned in the transverse direction. The idler beam passes through imaging lens L1 and then through a slit of width w. Collector lens L2 collects all light passing through the slit and focuses it onto detector D1.

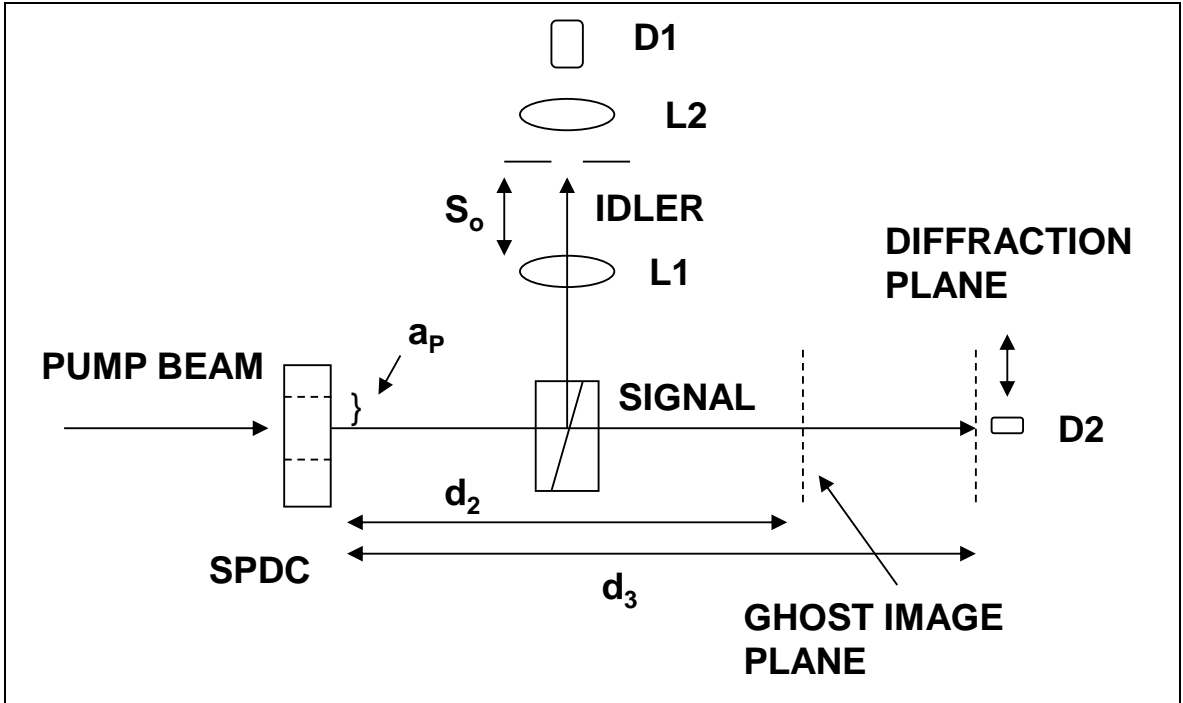

Fig. 1. Schematic diagram of entangled ghost imaging. Improve picture a bit.

The significance of the placement of the various elements is typically explained in terms of an unfolded geometry, proposed by Klyshko [15], as shown in Fig. 2, where we show relevant parameters for both ghost imaging and Popper diffraction.

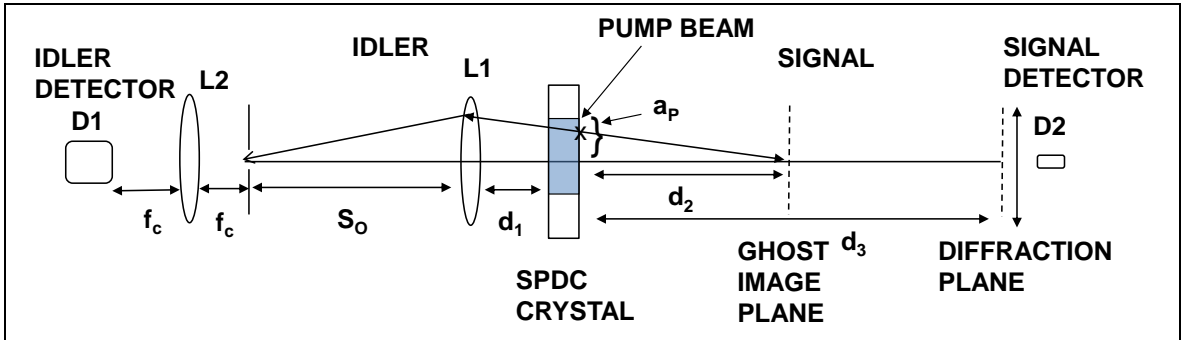

Fig. 2. Unfolded geometry for entangled ghost imaging.

In the unfolded geometry of Fig. 2, the ghost imaging properties are understood in terms of geometrical optics. The entangled signal-idler photons are assumed to be generated at a common point in the SPDC crystal. The signal-idler pair is entangled by the conservation properties of the SPDC process.

| | | |
|---|---|---|
| energy conservation | $\omega_s + \omega_i = \omega_P$ | (1a) |
| longitudinal phase matching | $k_{sz} + k_{iz} = k_{Pz}$ | (1b) |
| transverse phase matching | $\kappa_s + \kappa_i = \kappa_P$ | (1c) |
| phase entanglement | $\varphi_{\kappa i} + \varphi_{\kappa s} = \varphi_P$ | (1d) |

where $\omega_{s(i)(P)}$ is the frequency of the signal (idler) (pump), $k_{sz(iz)(P)}$ is the z component of the k-vector of the signal (idler) (pump) and $\kappa_{s(i)}$ is the transverse component of the k-vector of the signal (idler).

When the pump beam is collimated in the SPDC crystal, the range of $\kappa_P$ is small. Then, because of the entanglement in transverse k-vectors, the signal-idler pair forms a single "ray" in the geometrical optics picture. In this picture the plane containing the slit in the idler beam and the plane of the ghost image are conjugate image planes related by the thin lens equation

$$\frac{1}{S_o} + \frac{1}{S_i} = \frac{1}{f} \qquad (2)$$

where the quantities are as defined in Fig. 2, f is the focal length of L1, $S_i = d_1 + d_2$ and the magnification $M = (d_1 + d_2)/S_o$.

We will model the system in Fig. 1 in the Heisenberg picture. In this picture the field annihilation operators for the signal and idler, $a_s(x,z,t)$ and $a_i(x,z,t)$, are evolved from the source at the entrance face of the SPDC crystal to their respective detectors. The correlated counting rates at the detectors are formed as

$$CCR(x_{idet}, z_{idet}, x_{sdet}, z_{sdet}) = \\ < 0|a_i^\dagger(x_{idet}, z_{idet})a_s^\dagger(x_{sdet}, z_{sdet})a_s(x_{sdet}, z_{sdet})a_i(x_{idet}, z_{idet})|0>. \qquad (3)$$

where $x_{i(s)det}$ is the transverse position in the plane of the idler (signal) detector $z_{i(s)det}$.

The evolution of the operators through the SPDC crystal is described using the Heisenberg equation of motion

$$\frac{da}{dt} = [a, H] \qquad (4)$$

as discussed in [16] and [17]. The Hamiltonian is given by

$$H = H_o + i\eta \int dx \int dz\, A_P(x,z)\, a_s^\dagger(x,z)\, a_i^\dagger(x,z) \qquad (5)$$

where $H_o$ is the unperturbed Hamiltonian, $\eta$ describes the strength of the SPDC coupling and $A_P(x,z)$ is the amplitude of the pump. In our analysis we will assume that the pump beam has a Gaussian cross section with a dimension large enough so it does not diffract significantly within the crystal and therefore has the form

$$A_P(x,z) = A_P e^{-x^2/a_P^2} e^{ik_{Pz}z}. \qquad (6)$$

The operators for the signal and idler at the exit face of the SPDC crystal are

$$a_s(x_{os}, L) = \int d\kappa_s\, e^{i\kappa_s x_{os}} e^{i\varphi_{\kappa s}} a_s(\kappa_s, \varphi_s, \omega_s, L) \qquad (7a)$$

$$a_i(x_{oi}, L) = \int d\kappa_i \ e^{i\kappa_i x_{oi}} e^{i\varphi_{\kappa i}} a_i(\kappa_i, \varphi_i, \omega_i, L) . \tag{7b}$$

where $a_{s(i)}(\kappa_{s(i)}, \varphi_{s(i)}, \omega_{s(i)}, L)$ is the annihilation operator for a signal (idler) field mode with transverse wave vector $\kappa_{s(i)}$, phase $\varphi_{s(i)}$ and frequency $\omega_{s(i)}$. Using equations 5, 1b and 1c we obtain [see 16]

$$a_s(\kappa_s, \varphi_s, \omega_s, L) = a_{so} + i\eta \int d\kappa_P \ A_P(\kappa_P) a_i^\dagger(\kappa_P - \kappa_s) \text{sinc}(\Delta k_z L/2) e^{i\Delta k_z L/2} \tag{8a}$$
$$a_i(\kappa_i, \varphi_i, \omega_i, L) = a_{io} + i\eta \int d\kappa_P \ A_P(\kappa_P) a_s^\dagger(\kappa_P - \kappa_i) \text{sinc}(\Delta k_z L/2) e^{i\Delta k_z L/2} \tag{8b}$$

where $a_{s(i)o}$ is the free field operator in the absence of the SPDC interaction, $A_P(\kappa_P)$ is the pump amplitude with transverse wave vector $\kappa_P$ and $\Delta k_z$ is given by

$$\Delta k_z = \frac{\kappa_P^2}{2k_P} - \frac{\kappa_s^2}{2k_s} - \frac{\kappa_i^2}{2k_i} \tag{9}$$

The relation in equation 1d accounts for the phase entanglement of the signal and idler and produces a coherent propagation in the biphoton that is not present in the independent propagation of the signal and idler. We assume a single mode for the pump so that the phase of the pump, $\varphi_P$, is a constant.

After the SPDC crystal field operators will be propagated from position $(x_1, z_1)$ to $(x_2, z_2)$ by successive use of the Kirchoff diffraction integral in the Fresnel approximation [see 17]. For free space propagation the transport integral has the form [e. g. 18]

$$a(x_2, z_2) = e^{ik(z_2 - z_1)} \int dx_1 e^{i\pi(x_2 - x_1)^2/\lambda(z_2 - z_1)} a(x_1, z_1) \tag{10}$$

In our calculations we will assume that the process is effectively phase matched so that $\Delta k_z = 0$, and that the SPDC interaction is operated at degeneracy so that $\omega_s = \omega_i$.

The correlated counting rate of equation (3) for the signal in the diffraction plane $d_3$ and a spatially integrating idler detector in the focal plane of the collecting lens is

$$CCR(f_c, x_{3s}, d_3) = \int dx_{D1} \ |g(x_{D1}, f_c, x_{3s}, d_3)|^2. \tag{11}$$

The correlated counting rate for the spatially integrating idler detector in the focal plane of the collecting lens and the signal in the ghost image plane is given by

$$CCR(f_c, x_{2s}, d_2) = \int dx_{D1} \ |g(x_{D1}, f_c, x_{2s}, d_2)|^2. \tag{12}$$

In equations (11) and (12) $x_{D1}$ is the transverse position in the focal plane of the collector lens and the mode functions are given by

$$g(x_{D1}, f_c, x_{3s}, d_3) = \int_{-w/2}^{w/2} dx_{1i} \ e^{-\frac{i2\pi x_{1i} x_{D1}}{\lambda f_c}} f(x_{1i}, z_{1i}, x_{3s}, d_3) \tag{13a}$$

$$g(x_{D1}, f_c, x_{2s}, d_2) = \int_{-w/2}^{w/2} dx_{1i}\, e^{-\frac{i2\pi x_{1i} x_{D1}}{\lambda f_c}} f(x_{1i}, z_{1i}, x_{2s}, d_2) \quad (13b)$$

$$f(x_{1i}, z_{1i}, x_{3s}, d_3) = e^{i\pi x_{3s}^2/\lambda(d_3-d_2)} \int_{-b}^{b} dx_{2s}\, e^{\frac{i\pi x_{2s}^2}{\lambda(d_3-d_2)}} e^{-\frac{i2\pi x_{2s} x_{3s}}{\lambda(d_3-d_2)}} f(x_{1i}, z_{1i}, x_{2s}, d_2) \quad (14a)$$

$$f(x_{1i}, z_{1i}, x_{2s}, d_2) = e^{i\pi x_{2s}^2/\lambda d_2} \int dx_{os}\, e^{-\frac{x_{os}^2}{a_P^2}} e^{-\frac{i2\pi x_{os}[Mx_{1i}+x_{2s}]}{\lambda d_2}} \quad (14b)$$

where $x_{1i}$ is the transverse position in the plane of the idler slit $z_{1i}$. When a physical slit is placed in the signal at the ghost image plane, the limits on the integral in equation (14a) are set at $b = w/2$. When the slit is not present in the signal the limits on the integral in equation 14a are set at $b = \infty$.

In deriving equations 14a and 14b we integrated over lens L1 by stationary phase and kept the integral over the surface of the SPDC crystal $x_{os}$ because the effective limiting aperture stop in the geometry of Fig. 1 with the parameters of [2,3] is the diameter of the pump beam at the crystal and not the diameter of the imaging lens.

RESULTS

We consider first the results that are calculated with the parameters used in the experiments of [2] and [3]. The relevant parameters are $d_2 = 745$ mm, $S_o = 1000$ mm, M = 1, f = 500 mm, $d_3 = 1245$ mm. The diameter of the pump beam was stated in [2,3] to be 3 mm. We assume here that the pump beam has a Gaussian profile with a diameter $d_P = 2 a_P$ of 3 mm at the $1/e^2$ intensity point and a waist located at the crystal. The correlated counting rate for the signal as a function of transverse position in the diffraction plane ($d_3$) is shown in Fig 3 for a spatially integrating idler detector in the focal plane of the collecting lens L2 (equation 11). The solid curve (red) shows the signal distribution when no slit is located in the ghost image plane and the dotted curve (blue) shows the distribution when a slit of the same size as the one in the idler beam (160 μm) is placed in the signal at the ghost image plane. The idler slit is shown as the dashed curve (green). The width of the distribution with the physical slit in the signal at the ghost image plane is wider than the width of the distribution without the physical slit. This is the same behavior that was reported in [2,3], where the smaller width of the distribution without the slit was interpreted as demonstrating Popper's prediction.

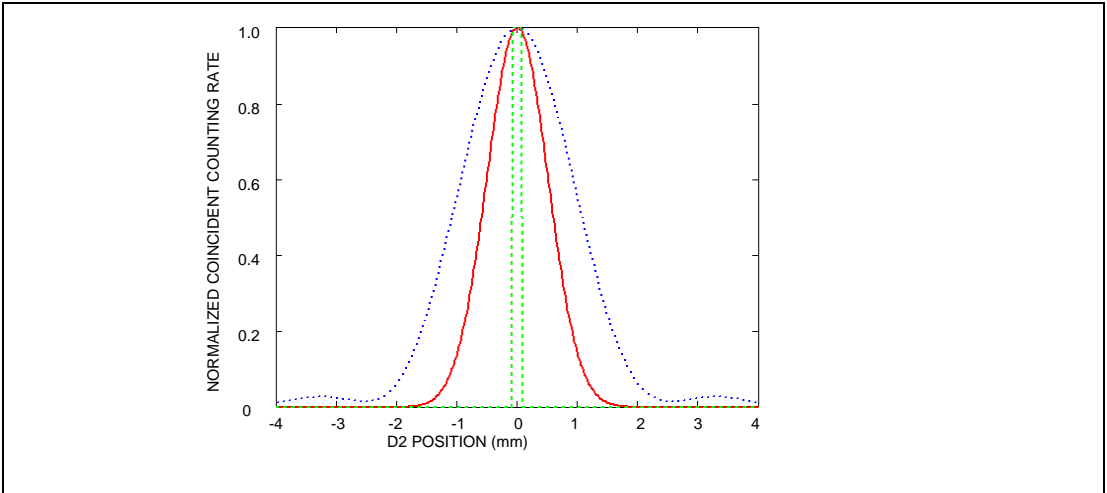

Fig. 3. Signal distribution in the diffraction plane with a spatially integrating idler detector
Red (solid) – no slit in signal FWHM = 1.17 mm
Blue (dotted) – slit in signal FWHM = 2.16 mm
Green (dashed) – idler slit profile

We will show here that the behavior shown in Fig. 3 with or without the slit is simply due to propagation of the mode distribution of the biphoton in the ghost plane as expected from conventional physical optics. No paradoxes or other contradictions are involved.

To illustrate this we consider the distribution of the correlated counting rate of the signal in the ghost image plane as shown in Fig. 4 for the integrating idler detector (red, solid curve) (equation 12) and for a point idler detector in the focal plane of the collecting lens L2. (blue, dotted curve). The correlated counting rate for the point idler detector is given by equation (11) with $x_{D1} = 0$

$$CCR(0, f_c, x_{2s}, d_2) = |g(0, f_c, x_{2s}, d_2)|^2 \qquad (15)$$

where $g(0, f_c, x_{2s}, d_2)$ is given by

$$g(0, f_c, x_{2s}, d_2) = e^{i\pi x_{2s}^2/\lambda d_2} \int_{-w/2}^{w/2} dx_{1i} \int dx_{os} e^{-\frac{x_{os}^2}{a_P^2}} e^{-\frac{i2\pi x_{os}[x_{1i}+x_{2s}]^2}{\lambda d_2}} . \qquad (16)$$

Also shown for comparison is the slit in the idler (green, dashed curve).

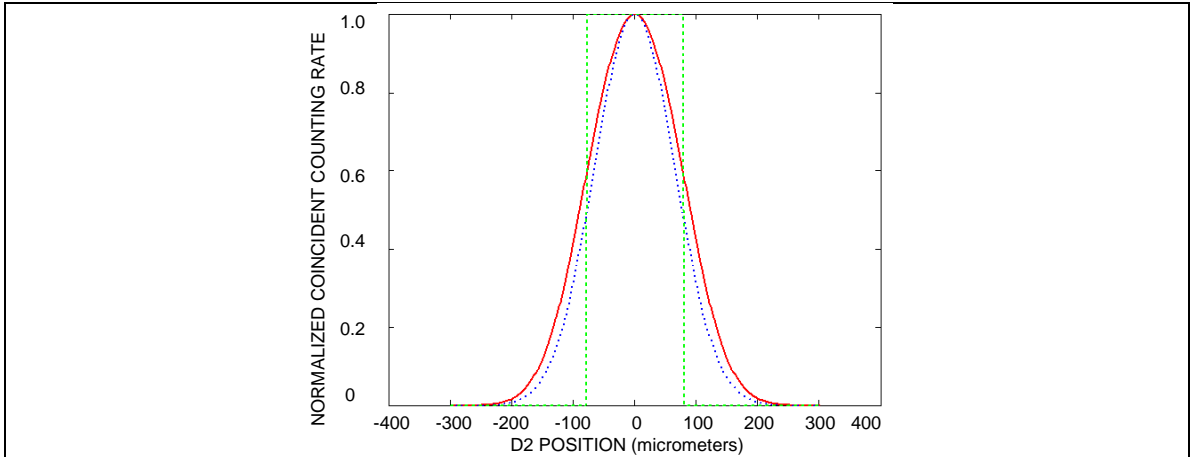

Fig 4. Signal distribution in the ghost image plane
Red (solid) – large detector  FWHM = 181 μm
Blue (dotted) – point  idler detector FWHM = 157 μm
Green (dashed) – idler slit profile

It is seen that the width of the signal distribution in the ghost image plane for either detector as characterized by the full width at half maximum (FWHM) is indeed about the same as the width of the idler slit as claimed in [2,3].  However, the mode profile of the signal in the ghost image plane is not a faithful reproduction of the slit as implied in [2,3].  Rather the signal profile is broadened in the wings to a dimension larger than the slit.  This is due simply to the limitation on the resolution of the ghost image imposed by the effective numerical aperture resulting from the finite pump beam diameter.  These conclusions are in agreement with the comments of Short [8], who gave a qualitative discussion of the effect of the pump beam diameter on the resolution of the ghost image and its subsequent propagation.

We confirm this explanation by increasing the effective numerical aperture by a factor 5 by setting $d_2 = 149$ mm.  The signal distributions for the integrating and point detectors in the ghost image plane are now as shown in Fig. 5.

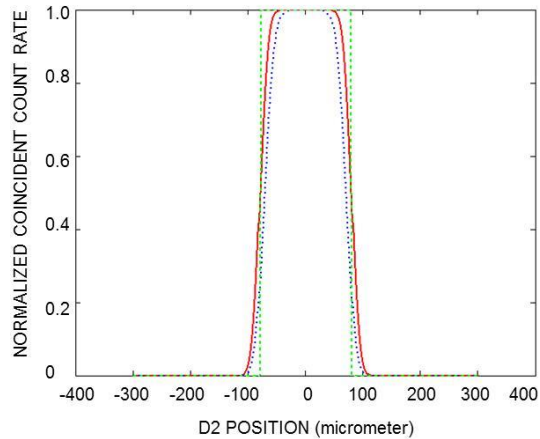

Fig. 5. Signal distributions in the ghost image plane with the large numerical aperture
Red (solid) – integrating idler detector FWHM = 80 μm
Blue (dotted) – point idler detector FWHM = 70 μm
Green (dashed) – idler slit profile

With the large numerical aperture the signal distribution in the ghost image plane forms a much more faithful reproduction of the idler slit with very little extension beyond the slit in the wings. The distributions of the signal in the diffraction plane with the larger numerical aperture are shown in Fig. 6 without (red, solid) and with (blue, dotted) the slit in the signal at the ghost image plane.

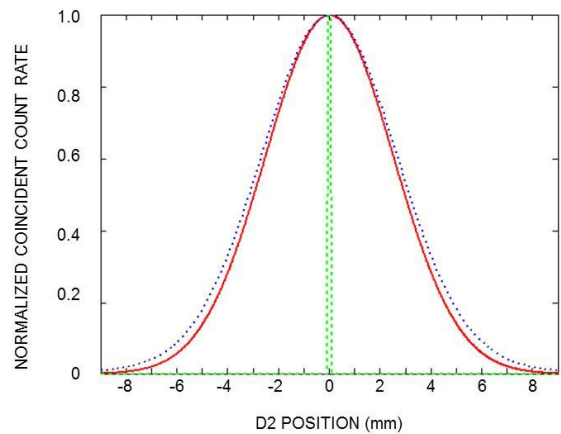

Fig. 6. Signal distribution in the diffraction plane with the large numerical aperture and the integrating idler detector
Red (solid) – no slit in ghost image plane FWHM = 5.9 mm
Blue (dotted) – slit in ghost image plane FWHM = 6.3 mm
Green (dashed) – idler slit profile

With the larger numerical aperture it can be seen that the presence of the physical slit in the signal at the ghost image plane has little to no effect on the width of the signal distribution in the diffraction plane.

We therefore conclude that the different widths of the distributions in the diffraction plane shown in Fig. 5 simply reflect the properties of optical propagation of the biphoton with and without a slit in the ghost image plane. The wider pattern observed in the diffraction plane when the slit is used in the signal at the ghost image plane results from the increased diffraction associated with the narrower, square distribution in the ghost image plane that is produced when the slit blocks the wings of the signal distribution. This behavior illustrates the importance of taking the mode profiles into account when dealing with problems involving optical propagation as has been pointed out by Menzel et al. in another context [19].

RELATION TO THE POPPER CONJECTURE

In order to relate these results to the Popper conjecture, we solve equation 11 analytically using a sequence of integrations by completing the square in the Gaussian forms. For the parameters of the system considered here, the result is

$$CCR(f_c, x_{3s}, d_3) = K e^{-\left(x_{3s}/\sigma_d\right)^2} \qquad (17)$$

where

$$\sigma_d = \frac{(d_3 - d_2)}{\sqrt{2} d_2} a_P \qquad (18)$$

and the normalizing constant $K = \int_{-w/2}^{w/2} dx_{1i} e^{(\pi a_P x_{1i}/\lambda d_2)^2}$ does not depend on $x_{3s}$.

Equations 17 and 18 show that the width of the signal distribution in the diffraction plane does not depend on either the width of the idler slit, w, or the wavelength $\lambda_s$ when an integrating idler detector is used, as in the experiments of [2,3]. Therefore, the width of the distribution in the diffraction plane does not arise from diffraction of an amplitude distribution in the ghost image plane. As a result the width of the signal distribution in the ghost image plane, $\Delta x$, and its width in the diffraction plane, $\Delta p$, are not Fourier transform pairs or quantum conjugates, and therefore are not subject to a Heisenberg-like restriction on the value of the product $\Delta x \Delta p$. In addition, the width of the signal distribution in the diffraction plane does not increase in inverse proportion to the width of the signal distribution in the ghost image plane in contradiction to the conjecture of Popper, but in agreement with all the other assessments of the conjecture.

There is a very simple geometric interpretation of equations 17 and 18 which is illustrated in Fig. 7.

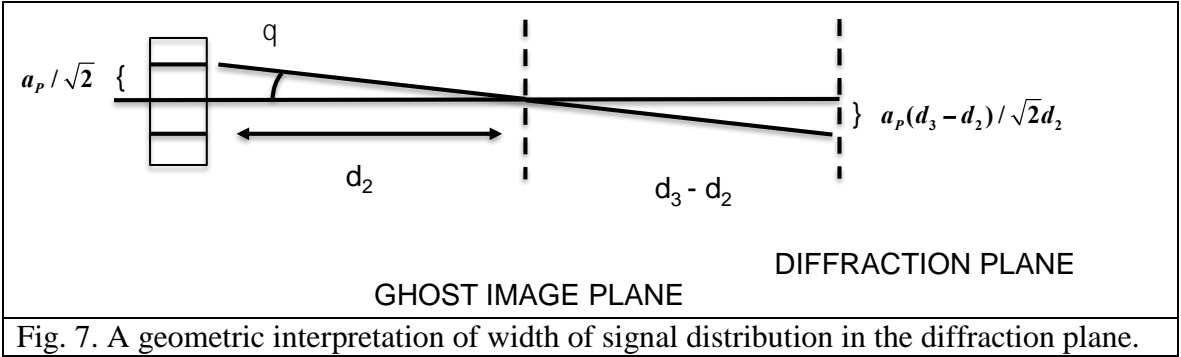

Fig. 7. A geometric interpretation of width of signal distribution in the diffraction plane.

The half-width of the distribution in the diffraction plane in equation 18 can be expressed as

$$\sigma_d = (d_3 - d_2)\theta \qquad (19)$$

where the angle $\theta$ is given by

$$\theta = \frac{a_P}{\sqrt{2}d_2} \qquad (20)$$

The angle $\theta$ in equation 20 is the effective numerical aperture of the optical system that determines both the resolution of the ghost image and the spread of the distribution in the diffraction plane.

A POTENTIAL REALIZATION OF THE POPPER CONJECTURE

We now consider a configuration in which the idler is detected with a point detector located at the center of the distribution in the focal plane of the collector lens L2. The correlated counting rate for the signal in the ghost image plane is given in equation 16. An analytic expression for $g(0, f_c, x_{2s}, d_2)$ can be obtained by completing the squares in the integrals

$$g(0, f_c, x_{2s}, d_2) = \text{erf}[(x_{2s} + w/2)(\pi a_P/\lambda d_2)] - \text{erf}[(x_{2s} - w/2)(\pi a_P/\lambda d_2)] \qquad (21)$$

where

$$\text{erf}(u) = \int_0^u dy\, e^{-y^2} \qquad (22)$$

The corresponding correlated counting rate is shown in Fig. 8 for the large numerical aperture.

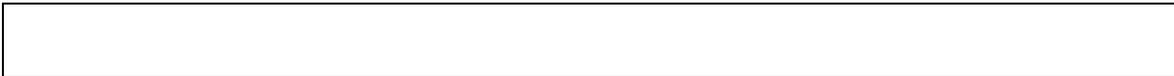

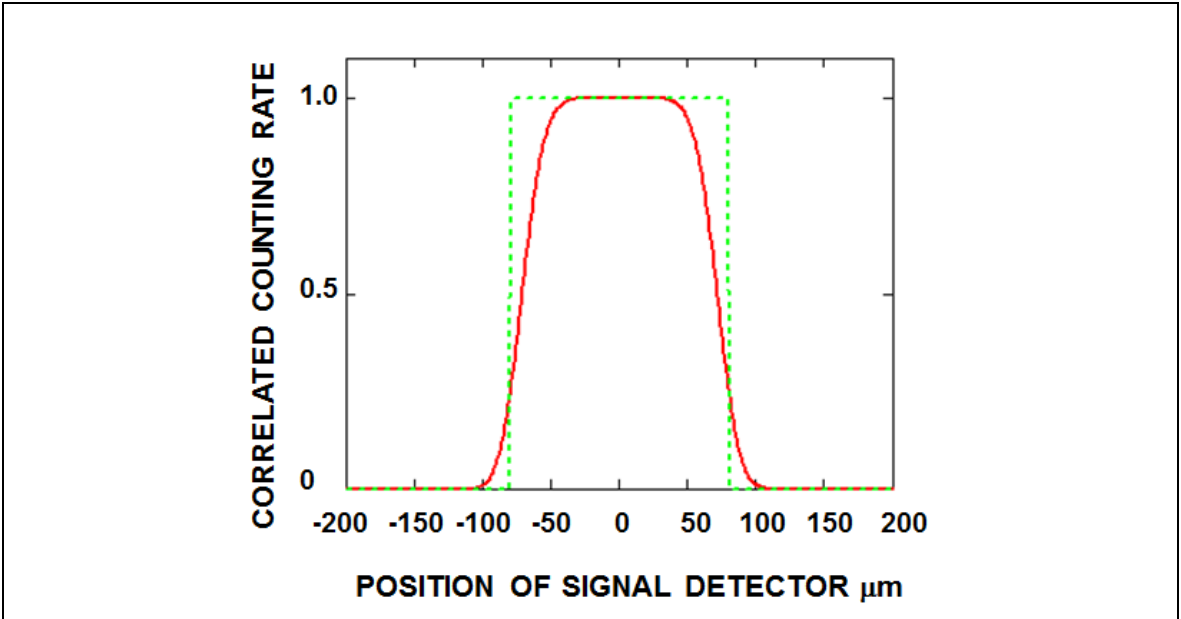

Fig. 8. Red (solid) – Analytic expression for the signal distribution in the ghost image plane for a point idler detector at the center of the collector lens focal plane. Green(dashed) – idler slit.

The distribution shows a reasonable representation of the idler slit with a FWHM of approximately 0.9w for the parameters considered here with the larger numerical aperture.

If the distance $(d_3 - d_2)$ is large enough to be in the Fraunhofer regime, the correlated counting rate distribution in the diffraction plane $d_3$ can be written as

$$CCR(0, f_c, x_{3s}, d_3) = \left| \int d\, x_{2s} e^{-i2\pi x_{2s} x_{3s}/\lambda(d_3-d_2)}\, g(0, f_c, x_{2s}, d_2) \right|^2 \qquad (24)$$

It can be seen that equation (24) has the form of a Fourier transform of the amplitude distribution of the signal distribution in the ghost image plane unlike the distribution for the integrating detector in equation (12). The signal distribution in equation (24) is shown in Fig. 9.

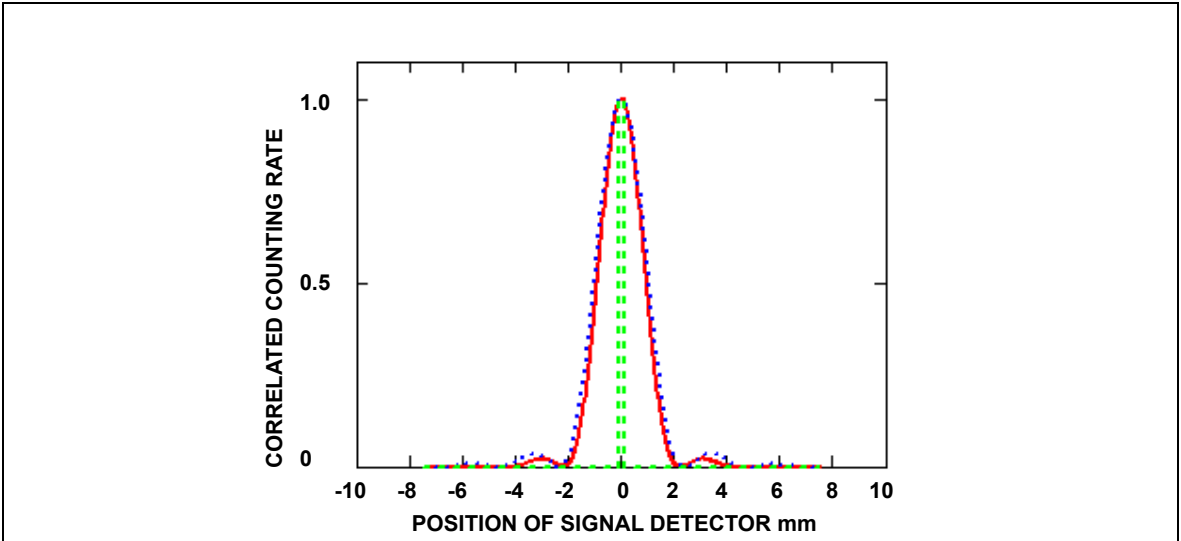

Fig. 9. Distribution of signal in the diffraction plane using a point idler detector in the ghost image plane and the large numerical aperture.
Red (solid) no slit in the signal at the ghost image plane
Blue (dotted) slit in the signal at the ghost image plane.
Green (dashed) idler slit.

It can be seen that with the point idler detector in the focal plane of the collecting lens L2 and the large numerical aperture, the signal distribution in the diffraction plane has the form of the Fourier transform of a slit and is considerably narrower than the distribution with the integrating detector shown in Fig. 6. From equations 22 and 24 it can be seen that the width of this distribution scales inversely with w, even if there is no physical slit in the signal at the ghost image plane.

SUMMARY AND DISCUSSION

We have analyzed the entangled ghost imaging interaction using full diffraction integrals and interpreted the results in terms of the Popper conjecture. Our results show that the increase of the width of the signal distribution in the diffraction plane that was previously reported in [2,3] when a physical slit was placed in the ghost image plane arises entirely because the physical slit blocks the wings of the signal mode profile in the ghost image plane. The effect disappears when the numerical aperture is increased to reduce the spread of the mode profile of the signal. Our results further demonstrate that, when an integrating idler detector is used, although the correlated signal distribution is confined in the ghost image plane, its width in the diffraction plane does not depend on either the width of the idler slit or the wavelength. Therefore the widths of the signal distribution in the ghost image plane and the diffraction plane are not conjugate pairs with a product that is restricted by a Heisenberg relation.

The behavior of the signal distributions in the ghost image and diffraction planes do not agree with the predicted behavior Popper attributed to quantum theory when the integrating idler detector is used. In his prediction, the second particle would spread after the virtual confinement plane at a rate that is inversely proportional to the width of its virtual confinement as dictated by the Heisenberg relation.  In our analysis the signal beam diverges after being virtually confined in the ghost image plane at a rate that is unrelated to the width of the slit in the idler and is much faster than the rate that would be imposed by a Heisenberg uncertainty principle related to the width of the slit. The results of our analysis are, however, similar to others that have been reported in analyses of the Popper conjecture using quantum particles.

Finally we identified a configuration that does reproduce the behavior that Popper attributed to quantum theory.  This configuration uses the ghost imaging configuration of [2,3] with a point idler detector at the center of the focal plane of the collector lens [equivalent to the point detector used in 11].  The point idler detector senses only those idler photons that are transmitted through the slit and are contained in a single Fourier component of the slit.  For this combination, the signal distribution in the ghost image plane shows a reasonably faithful reproduction of the idler slit with a width of 0.9 w for the large numerical aperture considered here.  The signal distribution in the diffraction plane is the Fourier transform of the amplitude of the distribution in the ghost image plane and its width increases in inverse proportion to the width of the signal in the ghost image plane even when no physical slit is placed in the signal.

It was this behavior that Popper rejected as being unphysical.  Popper cast the description of his experiment in terms of the role of observer knowledge.  In his viewpoint, the quantum theory predicted that the behavior of the second, unmeasured particle (the signal photon here) would be changed because of the knowledge about the transverse position of the first particle that is gained when it passes through the slit.  In particular the transverse position of the second particle would be confined when the observer learns of the position of the first particle, and the subsequent spread of the second particle (transverse momentum) would increase in inverse proportion to the dimension of the confinement because of that knowledge.

The new configuration using the point idler detector provides the fundamental test of quantum mechanics that Popper tried to produce with his original thought experiment. We offer an explanation of this predicted behavior that does not rely on the causative properties of observer knowledge.  Rather our analysis involves only the properties of the modes of the system.

Our explanation rests on the following premises

1. A photon is properly characterized as a unit excitation of a mode, or a superposition of modes, of the optical field.

2. The set of modes of the optical field that is available for excitation is determined by the configuration of the system under analysis.

3. The specific modes or combination of modes within that set in which excitations are observed are determined by the observer's choice of detector or combination of detectors.

Thus the role of observer knowledge in our explanation is not one of influencing the trajectory of unmeasured particles, but rather in identifying in which modes, or in which combination of modes, an excitation has been detected by virtue of a measurement of the first particle. The measured transverse distributions of the second particle are determined entirely by the properties of the excited modes.

With respect to the system shown in Fig. 1 or Fig. 2, the SPDC interaction produces a large number of modes for the signal and idler fields. Because of the conservation properties of the interaction the signal and idler modes are not independent of each other, but rather are entangled, forming modes of a signal-idler biphoton. When considered in isolation (that is, just the pump beam and the SPDC crystal) the SPDC interaction produces a large number of biphoton modes extending over a large region of space. The arrangement of the entangled ghost imaging geometry around the SPDC restricts the number and type of biphoton modes that will interact with suitably placed detectors. In particular, the presence of the idler slit in the geometry of Fig. 1 or Fig. 2 restricts the set of biphoton modes to those that have a non-zero complex amplitude in configuration space (here the transverse position $x_{1i}$ at the idler slit plane $z_{1i}$) in the region of the slit opening. The use of a point idler detector at the focal plane of the collector lens further restricts the set of modes that can be sensed to ones that comprise a single Fourier component of the slit. Note that this final mode is not necessarily a single mode of the original SPDC interaction. Rather it is a weighted combination of the original modes that satisfy the conditions imposed by the choice of the geometry and the detector.

When the point idler detector at the focal plane of the collector lens responds, it signifies that an excitation of the corresponding biphoton mode (or a set of those modes) has been detected. The biphoton mode in question has a profile described by equation 12. At the position of the ghost image plane, the amplitude of the signal part of the mode with the excitation has a nonzero value only in a region that extends approximately the width of the idler slit on either side of the z axis. If a signal detector were to be placed in this plane and its response were to be recorded in correlation with the idler detector, it would respond only when it was located in the region in which the mode amplitude is non zero. This does not mean that the signal photon was confined to this region because of the knowledge that the idler photon passed through the slit. Rather, it simply means that the mode sensed by the idler detection on that interaction had a nonzero amplitude in the region in question. If on the other hand, the idler detector were to be placed in the diffraction plane, it would respond according to the intensity distribution of the detected mode in the diffraction plane as shown in Fig. 9.

This relation between the signal distributions at the ghost image plane and the diffraction plane is exactly the behavior that Popper claimed quantum theory predicted. In his language, the extent of the unmeasured particle is confined because of the knowledge of the transverse position of its entangled twin, and the subsequent transverse spread of the unmeasured particle increases in inverse proportion to the size of the confinement. However, in our explanation, the particular pattern that is observed in the signal at the diffraction plane, which in this case corresponds to diffraction of the mode distribution at the ghost image plane, is not caused by the knowledge gained as a result of the idler measurement. Rather, it is simply the distribution in the signal diffraction plane of the biphoton mode that is sensed by the idler detector.

Similarly, if the point idler detector were replaced by a large area integrating detector, idler measurements would indicate excitations in a set of biphoton modes that cover a wider range of Fourier components of the slit. Each of these modes, corresponding to detection at a different position $xD1$ in the focal plane of the collector lens L2, has a nonzero signal amplitude in the ghost image plane in a transverse dimension equal to the slit width as shown in Fig. 10a. The signal distribution of each mode in the diffraction plane has the form of a diffraction pattern with a center position determined by the particular Fourier component of the slit that is associated with that mode as shown in Fig. 10b. Since the integrating idler detector senses the modes with all the Fourier components of the slit, the total signal distribution in the diffraction plane is the incoherent sum of the intensities of diffraction patterns from the individual modes associated with responses of point idler detectors at different positions in the collector lens focal plane as shown by the black (dash-dot) curve in Fig. 10b. Again, the observed signal distributions are not caused by the knowledge that is gained of the transverse position of the idler photons that pass through the slit through the slit. Rather the distributions simply reflect the appropriate combination of biphoton modes whose excitations are sensed by the integrating detector.

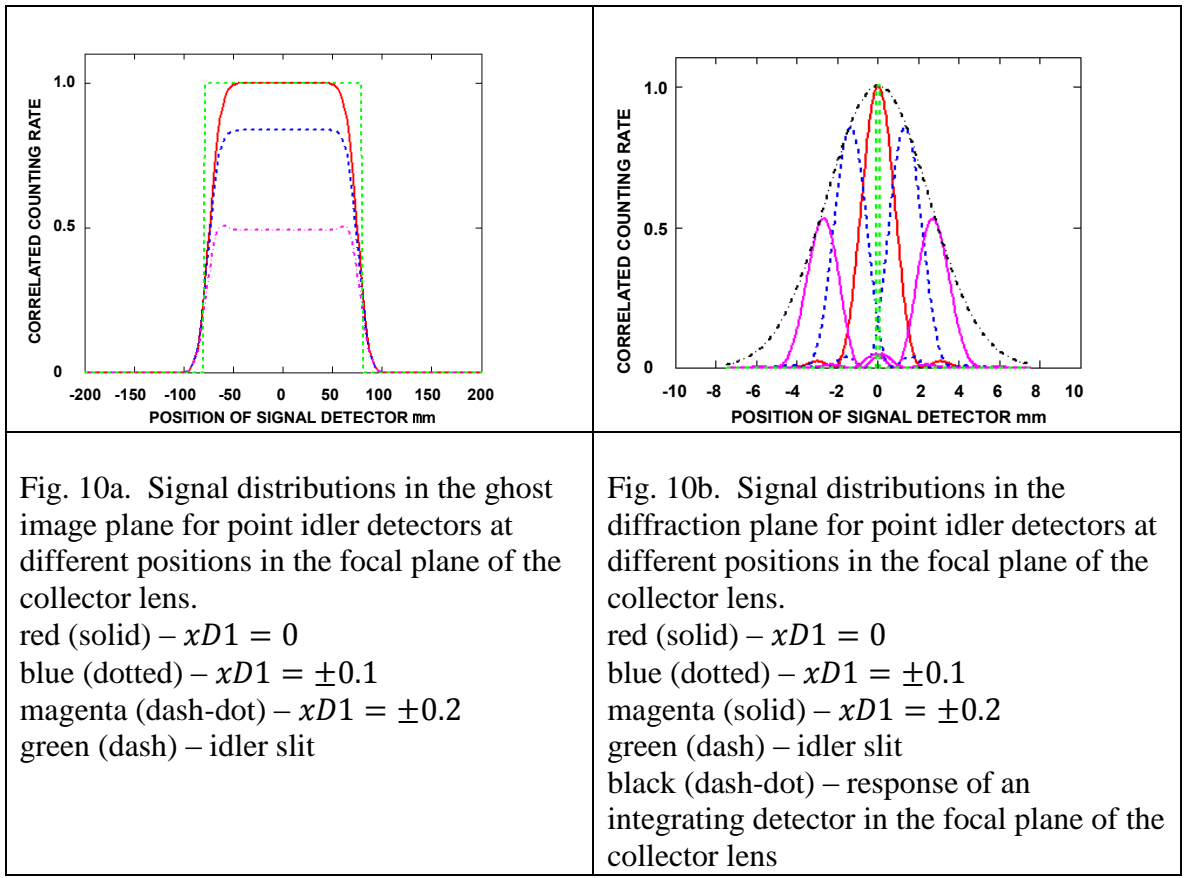

| Fig. 10a. Signal distributions in the ghost image plane for point idler detectors at different positions in the focal plane of the collector lens. <br> red (solid) – $xD1 = 0$ <br> blue (dotted) – $xD1 = \pm 0.1$ <br> magenta (dash-dot) – $xD1 = \pm 0.2$ <br> green (dash) – idler slit | Fig. 10b. Signal distributions in the diffraction plane for point idler detectors at different positions in the focal plane of the collector lens. <br> red (solid) – $xD1 = 0$ <br> blue (dotted) – $xD1 = \pm 0.1$ <br> magenta (solid) – $xD1 = \pm 0.2$ <br> green (dash) – idler slit <br> black (dash-dot) – response of an integrating detector in the focal plane of the collector lens |

This description is indicative of the behavior articulated by Marburger [20] that the properties we ascribe to quantum systems on the basis of particular measurements are strongly dependent on the nature of the measurements we decide to make.

Acknowledgements. This work was supported by the Office of Naval Research.REFERENCES

1. K. R. Popper, Zur Kritik der Ungenauigkeisrelationen, Die naturwissenschaften,22 Helft, **48,** 807 (1934); Popper KR. "Postscript to the Logic of Scientific Discovery" Vol. 3: Quantum Theory and the Schism in Physics. Totowa, New Jersey: Rowman and Littlefield, (1982).

2. Yoon-Ho Kim and Yanhua Shih, "Experimental Realization of Popper's Experiment: Violation of the Uncertainty Principle?", Foundations of Physics, **29**, No. 12, (1999).

3. Y. Shih and Y-H Kim, Fortschr. Phys **48**, 5-7, 463-471 (2000).

4. D. Bedford and F. Selleri, Lett . Nuovo Cimento **42**, 325 (1985).

5. A. Sudbery, Philosophy of Science **52**, 470 (1985)

6. M. J. Collett and R. Loudon, Nature **326**, 671 (1987).

7. W. M. Shields,"A Historical Survey of Sir Karl Popper's Contribution to Quantum Mechanics", Quanta DOI: 10.12743, Volume **1**, Issue 1, Page 1, (2012).

8. A. J. Short, " Popper's Experiment And Conditional Uncertainty Relations", Foundations of Physics Letters, **14**, No. 3, (2001).

9. Tabish Qureshi, "Understanding Popper's Experiment", American Journal of Physics **73**, 541 (2005); doi: 10.1119/1.1866098; Tabish Qureshi, "Analysis of Popper's Experiment and its Realization", Progress of Theoretical Physics, **127**(4), (2012).

10. Chris D. Richardson and Jonathan P. Dowling, "Popper's Thought Experiment Reinvestigated" International Journal Of Quantum Information **10**(3), 1250033 (2012).

11. D. V. Strekalov, A. V. Sergienko, D. N. Klyshko and Y. H. Shih, "Observation of Two-Photon "Ghost" Interference and Diffraction" Phys. Rev. Lett. **74** 3600 (1995).

12. Ming-Yang Zheng, Lian Chen, Feng Li, Ge Jin, "Ghost interference with spatially correlated photon resource" Optical Engineering **52**(5), 058002 (2013).

13. T. B. Pittman, Y. H. Shih, D. V. Strekalov, and A. V. Sergienko, Optical imaging by means of two-photon quantum entanglement, Physical Review A **52**(5) (1995).
18

14. Y. Shih, IEEE Journal Of Selected Topics In Quantum Electronics, **13**(4), July/August (2007).

15. D. N. Klyshko, "Photon and Nonlinear Optics", Gordon and Breach Science, New York, (1988).

16. Hong, C. K. and Mandel L.: Theory of parametric frequency down conversion of light. Phys. Rev. A **31**, 2409-2418 (1985).

17. Abouraddy, A. F., Saleh, B. E. A., Sergienko, A. V. and Teich, M. C.: Entangled-photon Fourier optics, JOSA B **19**, 1174-1184 (2001).

18. Goodman, J. W.; Introduction to Fourier Optics, chapter 5. McGraw-Hill, New York (1968).

19. R. Menzel, A. Heuer, D. Puhlmann, K. Dechoum, M. Hillery, M.J.A. Spahn and W.P. Schleich "A two-photon double-slit experiment", Journal Of Modern Optics **60**, 86-94 (2013). http://dx.doi.org/10.1080/09500340.2012.746400; Ralf Menzel, Dirk Puhlmann, Axel Heuer, and Wolfgang P. Schleich.

20. J. Marburger, "Constructing Reality", Cambridge University Press, New York (2011).